\documentclass[seceq]{ptptex}

\usepackage{graphicx}
\title{%        %You can use \\ for explicit line-break
Exotic Hadron in Pole-dominated QCD Sum Rules%
}

\author{%       %Use \scshape  for the family name
Toru Kojo %\textsc{kojo}
$^{1,}$\footnote{ e-mail address:
torujj@ruby.scphys.kyoto-u.ac.jp},  
Daisuke Jido, %\textsc{jido}
$^{2}$
and
Arata Hayashigaki %\textsc{haya}
$^{3}$
}

\inst{%         %Affiliation, neglected when [addenda] or [errata]
$^1$ Department of Physics, Kyoto University, Kyoto 606-8502, Japan \\
$^2$ Yukawa Institute for Theoretical Physics,
Kyoto University, Kyoto 606--8502, Japan \\
$^3$Institut f\"ur Theoretische Physik, J.W. Goethe Universit\"at,
          D-60438 Frankfurt am Main, Germany
}

\abst{
We study pentaquark $\Theta^{+} (I=0,J=1/2)$
in the QCD sum rules emphasizing that
we can not extract any properties of the pentaquark
outside of the Borel window.
To find the appropriate Borel window, 
we prepare a favorable set up of the correlators
and carry out the operator product expansion up to dimension 15
within factorization hypothesis.
Our procedures reduce the unwanted high energy contaminations
and enhance the low energy correlation.
In the Borel window, independent analyses for the chiral-even 
and odd sum rules give the consistent values of the $\Theta^+$ mass, 
$1.68\pm0.22$ GeV, and the residue. 
The parity is found to be {\it positive}.   
}

\begin{document}

\maketitle

\section{Introduction}
The experimental announcement for the
discovery of the pentaquark $\Theta^+(1540)$ \cite{leps}
triggered tremendous amount of theoretical and
experimental works on the exotic states.
Although the existence of such exotic states is still
not so obvious,
the exotics provide a good opportunity
to get the deeper insight of the hadron structures
and their connection to QCD.
One of approaches from QCD to exotics 
is the QCD sum rule (QSR)
\cite{shif},
which relates informations of  
QCD to the hadronic properties through the correlator analysis 
for the interpolating fields of hadrons.
The Borel transformed sum rules with
the simplest pole + continuum parametrization are given as
($i=0,\ 1$ correspond to the chiral even and odd part, respectively)
\begin{eqnarray}
\hat{L}_M\Pi^{(ope)}_i(-Q^2) = 
\lambda_i^2 e^{-m^2/M^2}
+\int_{s_{th}}^{\infty}\! ds\ e^{-s/M^2} 
\frac{1}{\pi} {\rm Im}\Pi^{(ope)}_i(s),
\end{eqnarray}
where the relation $\pm m\lambda_{0}^2 =\lambda_{1}^2$ 
holds due to the spinor structure
and the relative sign of the residues $\lambda_i^2$
represents with the parity of the resonance state.
Using these sum rules, we can derive the approximated expressions of the
mass and residue as a function of $M$ and $s_{th}$.
To extract properties of the low energy excitations with good accuracy,
we need to treat sum rules in the 
appropriate $M^2$-region, i.e., {\it Borel window},
where the low energy correlation is large enough compared to the
contaminations from high energy components 
which have no relations with properties of low-lying resonances. 
The setting the Borel window is the most important step in QSR and,
only within this window, 
we can evaluate the physical quantities. 

In the exotic cases, as reported in Ref. 3),
it is extremely difficult to find the appropriate
Borel window in contrast to the usual meson and baryon cases.
This is because the OPE convergence is very slow 
and the unwanted high energy components
dominate the spectral integral.
In addition,
we often encounter the {\it artificial stability} of the 
physical quantities against $M^2$-variation.
This is the case that physical quantities depend too strongly on the 
threshold parameter $s_{th}$ and not on 
the low energy correlations which we want to extract.
To attack these serious problems common to the
exotics,
we propose a new approach and apply it to the $\Theta^+$,
assuming its quantum number as $I=0,\ J=1/2$,
as one example of the exotics \cite{kojo}.

\section{OPE and favorable set up of the correlators}
To find the Borel window, it is necessary to 
increase low energy informations in the spectral function
efficiently and, at the same time,
reduce high energy contaminations.
For these purposes, we take the following treatments.

Inclusion of the higher dimension terms of OPE 
is particularly important because
they strongly reflect 
the low energy dynamics.
For example, in the case of the sum rules for $\rho$ and $A_1$ mesons,
the dim.0 and 4 terms are the same due to the chiral symmetry
realized in the high energy, 
and the splitting of the masses is explained only after the inclusion
of dim.6 terms, $\langle \bar{q}q \rangle^2$,
which appear due to the chiral symmetry breaking.
From these observations,
we perform the OPE calculation up to dim.15
within factorization hypothesis
both for taking into account the low energy correlations
and for the confirmation of good OPE convergence.
 
As later shown,
simple inclusion of the low energy correlations through the higher dimension
terms is found to be not sufficient to find the Borel window 
because high energy contaminations are too large in the 
QSR for the exotics.
To reduce the high energy contaminations,
we take the difference between correlators for two interpolating fields with 
{\it similar structure but different chirality} each other, i.e.,
\begin{eqnarray}
\lefteqn{ \hat{L}_M \bigg\{i\int d^4x \, e^{iq\cdot x}
\left\langle 0 \left|T [P(x)\bar{P}(0) 
- t\, S(x)\bar{S}(0)] \right|0 \right\rangle \bigg\} } 
\nonumber\\
&=& \int_0^{\infty}\!ds\ e^{-s/M^2}
\bigg\{ {\rm Im}[\Pi_0^P(s) - t\, \Pi_0^S(s)]\, \hat{q}
+{\rm Im}[\Pi_1^P(s) - t\,\Pi_1^S(s)] \bigg\}, 
\end{eqnarray}
where $\Pi_0$, $\Pi_1$ correspond to the chiral even and odd part
respectively, and 
\begin{eqnarray}
&&P = \epsilon^{abc} \epsilon^{def} \epsilon^{cfg} \{u^T_a C d_b\}
\{u^T_d C \gamma_{\mu} \gamma_5 d_e\}
 \gamma^{\mu} C \overline{s}^T_g,\\
&&S =  \epsilon^{abc}\epsilon^{def} \epsilon^{cfg} \{u^T_a C \gamma_5 d_b\}
\{u^T_d C \gamma_{\mu} \gamma_5 d_e\}
 \gamma^{\mu} C \overline{s}^T_g.
\label{eqn:current}
\end{eqnarray}
Here the only difference in these interpolating fields 
is that the first diquark structures have the opposite chirality.  

Let us first explain in the case of the chiral even part.
Since they show the same behavior in high energy due to the chiral
symmetry, 
after the subtraction of two correlators ($t=1$ case),
the irrelevant high energy contributions are expected to be canceled out
in the similar way as the Weinberg sum rules\cite{wein}.
In terms of OPE, this cancellation corresponds to 
the cancellation of the lower dimension terms.
It is not a priori evident 
whether the low energy correlations
remain enough even after the subtraction because
the low energy contribution could also cancel out.
Our Borel analysis, however, reveals that, in the case of $t=1$,
the large low energy correlation remains enough 
even after the subtraction.
As a result, we can find the Borel window in the relatively
large $M^2$-region.

On the other hand, for the chiral odd part, 
the subtraction procedure corresponding to $t=1$ case 
does not lead the cancellation of the high energy components because 
chiral odd part is constructed of the chiral symmetry breaking terms.
However, in the case of $t=1$,
the OPE convergence is found to be very good and then
we can find the Borel window in the small $M^2$-region
where high energy contaminations are suppressed due to 
the Borel factor $e^{-s/M^2}$ in the integral of the 
spectral function.
 
\section{Borel analysis for mass and residue}
Here we explain our criterion for the Borel window.
The lower bound of the Borel window is given
so that the highest-dimensional terms in the truncated OPE are 
less than 10\% of its whole OPE,
while the upper bound is determined by the region where
the absolute value of the pole contribution
is larger than the absolute value of the integrated spectral function
in the region $s \ge s_{th}$.
Note that the 50\% pole contribution in our criterion is extremely 
large in comparison with any prior pentaquark sum rules, 
where the pole contributions are not more than 20\% \cite{nari}.
\begin{figure}[b]
\vspace{-0.3cm}
\begin{center}
\includegraphics[width=12.0cm, height=4.5cm]{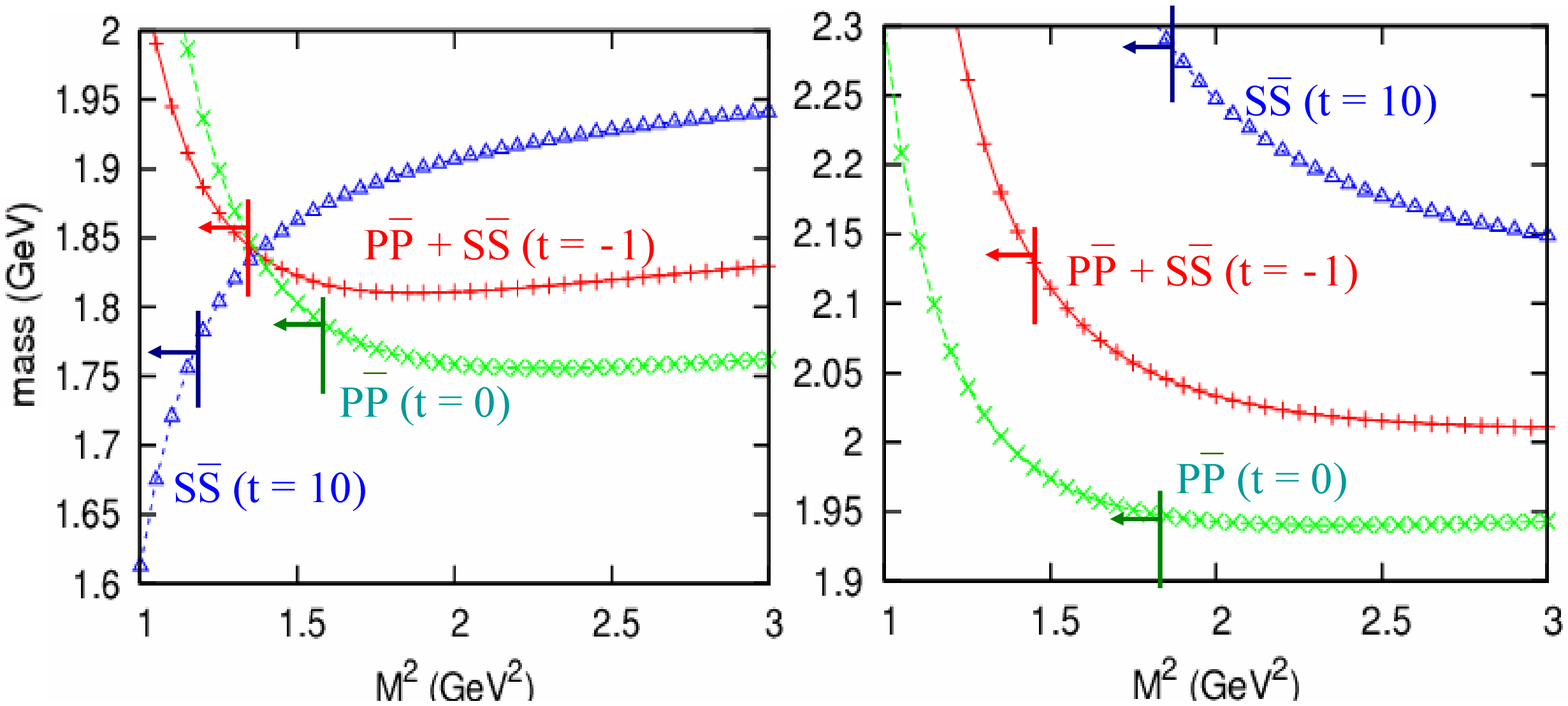}
\caption{\label{fig:cont1}
 The behavior of the mass
as a function of $M^2$ for  $t= -1,\ 0,\ 10$. The left arrows represent
the upper bound of the Borel window.
In the smaller $M^2$-region than the upper bound,
we can not find stable region of the mass. 
The stabilities above the upper bound are simply artifacts
which often appear in QSR.}
\end{center}
\vspace{-0.4cm}
\end{figure}

To recognize the problems in the case of QSR for exotics,
let us see Fig.{\ref{fig:cont1}} for $M^2$-dependence of the mass 
in the cases of $t= -1,\ 0,\ 10$
corresponding to $P\bar{P}+S\bar{S}$, $P\bar{P}$, $S\bar{S}$ 
cases respectively. 
The threshold value is fixed to typical value $\sqrt{s_{th}}=2.2 $ GeV.
In these cases, we fail to find stabilities of the mass
in the $M^2$-region lower than the upper bound of the Borel window.
The stabilities above the upper bound are simply artifacts
which often appear in QSR.
Fig.{\ref{fig:cont1}} shows that
typical mass of $P\bar{P}$ case is much smaller than that of $S\bar{S}$,
and then we can expect 
that the low energy correlation of $P\bar{P}$ is much larger
than that of $S\bar{S}$.
This observation leads that even after the subtraction
$P\bar{P}-S\bar{S}$ ($t=1$ case),
the low energy correlation can remain enough.

Now we see the case of around $t=1$.
We tune the value of $t$ around $t=1$ to get the widest Borel window.
As expected, for even part ($t=0.9$), 
the high energy contaminations are canceled out due to
chiral symmetry and
we find the wide Borel window in the relatively large $M^2$-region.
On the other hand, for odd part ($t=1.1$), 
thanks to the good OPE convergence, we also find
the wide Borel window in the small $M^2$-region.
The threshold values are taken to make the physical quantities
most stable in the Borel window.

The best stability is achieved with $\sqrt{s_{th}}= 2.2$ GeV (even) 
and $2.1$ GeV (odd), giving $m_{\Theta^+}= 1.64$ 
GeV (even) and $1.72$ GeV (odd) respectively.
\begin{figure}[t]
\begin{center}
\includegraphics[width=12.0cm, height=4.5cm]{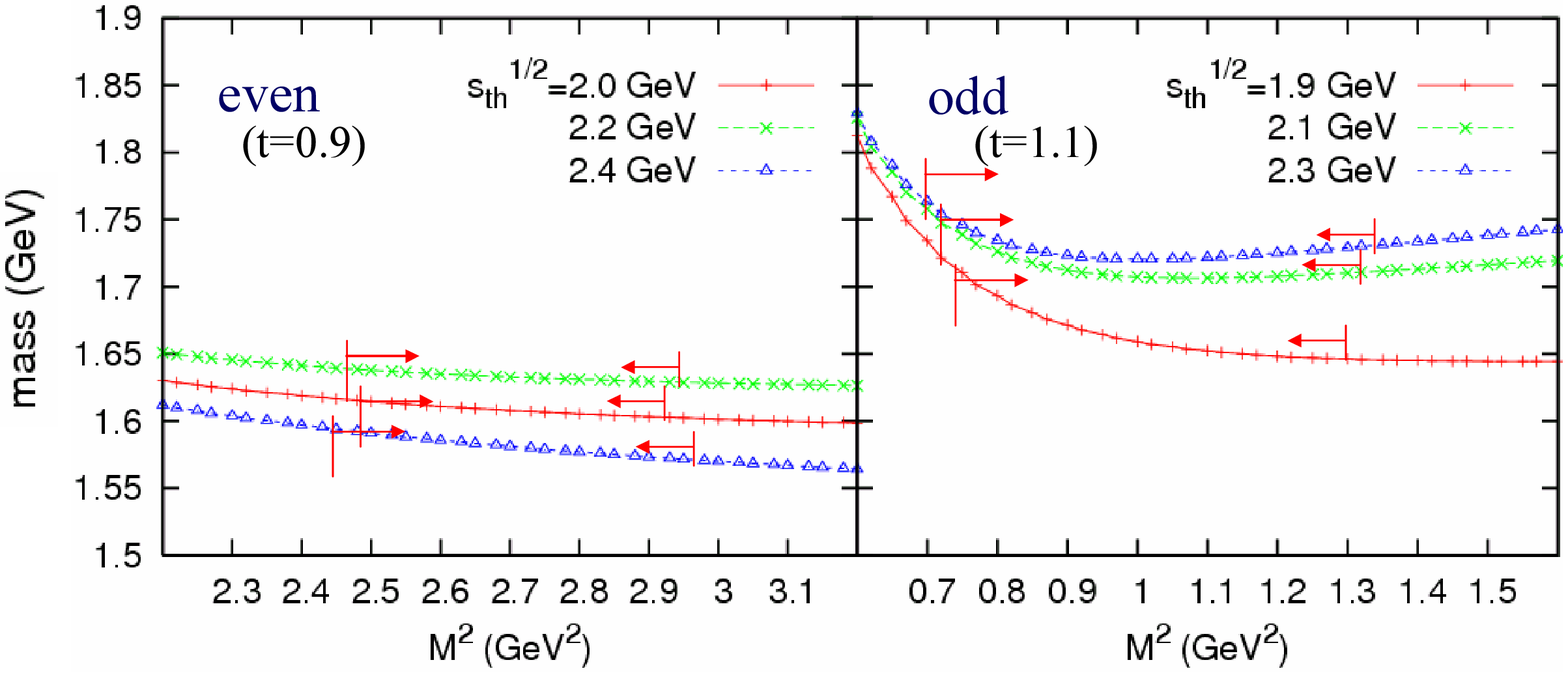}
\caption{\label{fig:cont2}
 The behavior of the mass
as a function of $M^2$. The left (right) arrows represent
the upper (lower) bound of the Borel window. 
We succeed to find the Borel window and stabilities of mass.}
\end{center}
\vspace{-0.5cm}
\end{figure}
The values of the residue are also obtained from the chiral even 
and odd sum rules as 
$\lambda_0^2 = (3.0 \pm 0.1)\times 10^{-9}$ GeV$^{12}$ 
and $\lambda_1^2/m_{\Theta^+} = (3.4 \pm 0.2) \times 10^{-9}$ GeV$^{12}$.
It is remarkable that these numbers are quite similar with
the close $t$. 
This implies that our analysis investigates consistently the same state
in the two independent sum rules. 
Note that from the relative sign of the residues,
we assign {\it positive} parity to the observed $\Theta^+$ state. 

In conclusion, we perform the Borel analysis for $\Theta^+$
with special setup of the correlators in order to find
the Borel window. Within uncertainties of the condensate value,
independent analyses for the chiral-even and odd sum rules
give the consistent values of the $\Theta^+$ mass, 
$1.68\pm0.22$ GeV, and the residue. 
The parity is found to be {\it positive}.

\section*{Acknowledgements}
We thank Profs. M. Oka, A. Hosaka and S.H. Lee for useful discussions
about QSR for the exotics during the YKIS2006 on "New Frontiers on QCD"
held at the Yukawa Institute for Theoretical Physics.
This work is supported in part by the Grant for Scientific Research
(No.18042001) and by Grant-in-Aid for the 21st Century
COE "Center for Diversity and Universality in Physics" from the Ministry of
Education, Culture, Sports, Science and Technology (MEXT) of Japan.

\end{document}